\documentclass[12pt]{spieman}  
\usepackage{amsmath,amsfonts,amssymb}
\usepackage{graphicx}
\usepackage{setspace}
\usepackage{tocloft}
\usepackage{lineno}
\usepackage{colortbl}
\title{Contour-Constrained Deformable Registration with Parameter Characterization for Head and Neck Surgical Guidance}

\author[a]{Qingyun Yang}
\author[b]{Jon S. Heiselman}
\author[a]{Ayberk Acar}
\author[b]{Morgan J. Ringel}
\author[b]{Michael I. Miga}
\author[a]{Matthieu Chabanas}
\author[b,c]{Michael C. Topf}
\author[a,b*]{Jie Ying Wu}
\affil[a]{Vanderbilt University, Department of Computer Science, 2301 Vanderbilt Pl, Nashville, United States}
\affil[b]{Vanderbilt University, Department of Biomedical Engineering, 2301 Vanderbilt Pl, Nashville, United States}
\affil[c]{Vanderbilt University Medical Center, Department of Otolaryngology-Head and Neck Surgery, 1211 Medical Center Drive, Nashville, United States}

\cftpagenumbersoff{figure}
\cftpagenumbersoff{table} 
\begin{document}

    \maketitle

\begin{abstract}
\textbf{Purpose:} With approximately 890,000 annual new cases globally, head and neck squamous cell carcinoma has one of the highest recurrence rates among solid malignancies. Although frozen section analysis is the standard of care for intraoperative margin assessment, accurately relocating detected positive margins on the resection bed remains challenging due to imprecise alignment between pathologic margins on resected specimens and the resection bed, compounded by post-resection mucosal tissue shrinkage. To address these challenges, we present a biomechanics-driven deformable registration framework that corrects post-resection tissue deformation to provide intraoperative guidance. \textbf{Approach:} Our approach registers three-dimensional specimen meshes to intraoperative resection bed point clouds using a deformable registration approach based on regularized Kelvinlet basis functions. The registration matches surface point clouds, fiducial landmarks, and boundary contour constraints that directly penalize perpendicular distance-to-agreement between specimen and resection bed boundaries. 
\textbf{Results:} Across nine head and neck specimens from three anatomical sites (skin, buccal mucosa, and tongue), the overall mean target registration error was $11.11 \pm 4.07$ mm using rigid registration, which decreased to $8.20 \pm 2.68$ mm (26.19\% reduction) using deformable registration without contour constraint. The proposed contour-constrained deformable registration further reduced the error to $5.62 \pm 2.28$ mm, representing a 49.41\% reduction relative to rigid registration. 
\textbf{Conclusions:} We observed the largest reduction in the most clinically challenging tongue specimens. We also performed a systematic two-stage parameter search to characterize how the relative importance of surface alignment, fiducial correspondences, contour constraint, and strain energy regularization varies across tissue types. The parameter search revealed that contour weighting dominates registration accuracy for tissue types with large lateral deformation, while the algorithm operates over a broad range of parameter combinations when specimens are pooled across tissue types. Together, these results provide a foundation for parameter selection in deformable registration of head and neck specimens.
\end{abstract}


{\noindent \footnotesize\textbf{*}Jie Ying Wu,  \linkable{JieYing.Wu@vanderbilt.edu} }

\begin{spacing}{2}   

\section{Introduction}
\label{sect:introduction}  
Head and neck squamous cell carcinoma (HNSCC) accounts for 890,000 new cases annually worldwide~\cite{barsouk2023epidemiology}. Surgical resection is one of the primary treatments for HNSCC~\cite{teixeira2025outcomes}. However, positive surgical margins are identified in up to 38.7\% of HNSCC resections, and recurrence rates reach 46.8\%~\cite{matsuo2024interval}. To promote local control~\cite{pierik2021resection}, frozen section analysis is commonly used to assess margin status intraoperatively. When positive or close margins are identified, it is critical for the surgeon to accurately relocate these margins based on the verbal description of frozen section diagnosis back onto the resection bed to establish the area of re-resection~\cite{nayanar2019frozen}. However, accurately relocating positive margins remains challenging due to the intricate three-dimensional (3D) anatomy in the head and neck and large deformations contributing to the persistently high rates of positive margins in head and neck surgeries.

Previous methods for establishing correspondence between the specimen and resection bed, such as surgical clips or paired tags~\cite{dzhugashvili2010surgical, van2019relocation}, do not provide dynamic spatial visualization of tumor and resection margin boundaries during surgery. With advancements in 3D scanning technology, the integration of virtual 3D models into pathology reports has helped to partially alleviate this limitation by enhancing intraoperative communication. However, inconsistencies in specimen orientation and positioning can still make these models imprecise and challenging to interpret, even for experienced surgeons and pathologists~\cite{miller2024far}.

Recently, head-mounted display augmented reality (AR) has gained traction in surgical applications~\cite{prasad2024more}. We have previously developed an AR guidance system to allow surgeons to overlay 3D holographic reconstructions of medical data and manipulate the hologram through gestures or voice commands~\cite{prasad2023augmented,tong2024development}. The hologram provides visual guidance to portray the orientation with which the extracted specimen corresponds to the resection bed. Additionally, the hologram can be annotated to show pathology margin findings and thereby improve the accuracy of margin relocation at intraoperative sites of residual tumor. Nevertheless, direct rigid registration of the specimen to the resection bed is still challenging due to intraprocedural soft-tissue deformations and mucosal shrinkage from cautery, dehydration, and other sources. Prior characterizations have demonstrated that mucosal specimens can shrink by 10-47\% post-resection~\cite{necker2023virtual}, which compromises the accuracy of these approaches. 

To address the challenge of tissue deformation, our prior work~\cite{Yang2025Nonrigid} proposed a nonrigid registration method for intraoperative specimen-to-resection bed alignment. A structured light 3D scanner is used to capture a 3D mesh of the surgical specimen, while an RGBD camera is used to acquire the point cloud of the resection bed. By registering the specimen mesh to the resection bed point cloud, the prior work deforms the specimen mesh to enhance alignment between specimen and resection bed. This nonrigid registration framework optimizes three key objective terms: (1) a surface alignment constraint that permits tangential sliding while enforcing normal correspondence between the specimen surface and resection bed points, (2) sparse fiducial point correspondences that enforce exact positional matches, and (3) a biomechanical regularization term based on strain energy that ensures physically plausible deformations. However, this prior work exhibited an important limitation: while the deformed specimen mesh achieved good surface alignment with the resection bed point cloud, the deformed mesh's boundary often extended beyond the boundaries of the resection bed. This occurred because the surface alignment constraint enforces point-to-surface correspondence without explicitly constraining the specimen's outer boundary to conform to the boundary geometry of the resection bed. The sparse fiducial points (typically four per specimen) provided insufficient constraints to fully determine the boundary shape, allowing the deformed mesh to adopt configurations that minimize surface distance while violating boundary correspondence.

As an extension of our prior work, this paper introduces two primary contributions. First, we propose the use of specimen boundary contour constraint that explicitly enforce alignment through normal-direction distances between resected specimen and resection bed boundaries, addressing the challenge of lateral boundary displacement that point-to-surface and point-to-point correspondences cannot sufficiently constrain. Second, we performed a systematic two-stage parameter search to characterize the relative importance of surface alignment, fiducial correspondence, and boundary contour constraint towards algorithm performance under varying strain energy regularization and across tissue types. We validated our approach on nine resection specimens across skin, buccal, and tongue tissue, with results reported across tissue types to reflect differences in post-resection deformation characteristics.

\section{Methods}
\label{sec:methods}

\begin{figure}[h!]
  \centering
  \includegraphics[width=\textwidth]{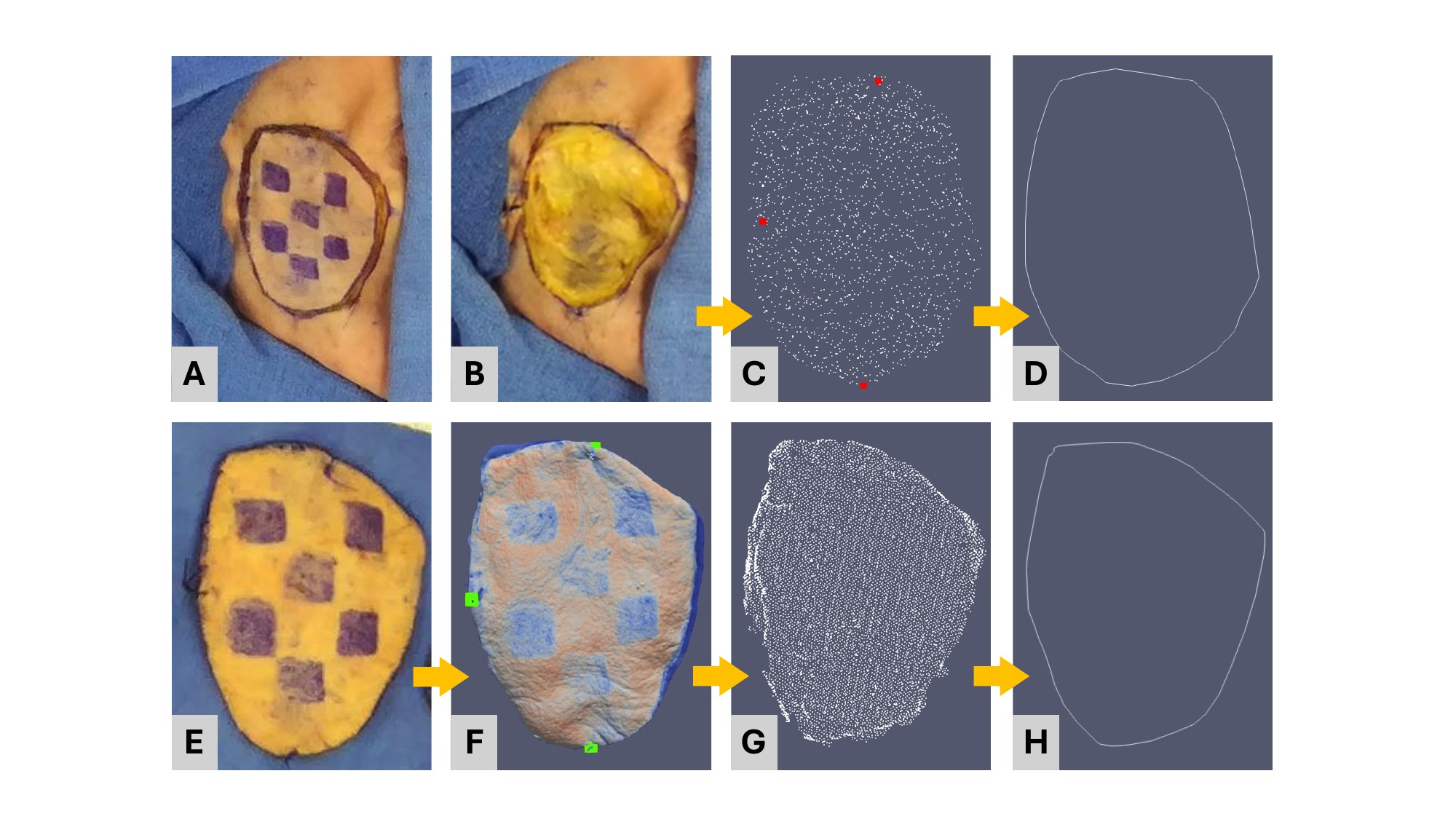}
  \caption{Data collection and pre-processing of a resected skin specimen. (A) Marked resection plan. (B) Resection bed. (C) Resection bed point cloud data with suture points marked in red. (D) Contour polyline extracted from resection bed point cloud. (E) The resected specimen with 3 suture points. (F) 3D specimen model with suture points marked in green. (G) Posterior surface point cloud of 3D specimen model. (H) Contour polyline extracted from the posterior surface point cloud}
  \label{fig:DataCollected}
\end{figure}

Figure~\ref{fig:DataCollected} provides an overview of our data collection and preprocessing workflow for a representative skin specimen. We begin with surgical planning and incision, during which we suture fiducial markers (Figure~\ref{fig:DataCollected} (A)). Following resection, we scan the resected specimen (Figure~\ref{fig:DataCollected} (E)) to obtain a 3D mesh (Figure~\ref{fig:DataCollected} (F)), and we capture the resection bed (Figure~\ref{fig:DataCollected} (B)) using an RGBD camera to generate a corresponding point cloud (Figure~\ref{fig:DataCollected} (C)). We then extract boundary contour polylines from both the specimen and the resection bed (Figure~\ref{fig:DataCollected} (D) and (H)). These steps produce the inputs required for deformable registration.

Our deformable registration framework corrects post-resection tissue deformation by aligning the 3D mesh of the resected specimen (Figure~\ref{fig:DataCollected} (F)) with the point cloud of the resection bed (Figure~\ref{fig:DataCollected} (C)). We achieve this by matching three types of geometric features. First, we match the posterior surface of the 3D specimen mesh (Figure~\ref{fig:DataCollected} (G)), referring to the tissue interface that was in contact with the underlying anatomy prior to resection and excluding the exterior surface, against its resection bed surface (Figure~\ref{fig:DataCollected} (C)). Second, we match the fiducial points on the resected specimen (green points in Figure~\ref{fig:DataCollected} (F)) against their corresponding fiducial points on the resection bed (red points in Figure~\ref{fig:DataCollected} (C)). These paired fiducial points are sutures placed at corresponding locations on the specimen and resection bed prior to resection, providing exact correspondence constraints. Third, we align the boundary contour line of the specimen (Figure~\ref{fig:DataCollected} (H)) with the boundary contour line of its resection bed (Figure~\ref{fig:DataCollected} (D)). Lastly, we incorporate a strain energy regularization term to penalize biomechanically implausible deformations.

The following subsections detail our data processing, including automated contour extraction, the deformable registration framework, parameter search, and experimental setup.

\subsection{Data Pre-processing}
\subsubsection{Surface Extraction}
We performed several preprocessing steps to prepare the data for deformation correction. For the surface constraint, we cropped and down-sampled the geometric data captured by the RGBD camera to obtain the point cloud of the resection bed (Figure~\ref{fig:DataCollected} (C)). We then extracted the vertices on the posterior surface from the specimen mesh acquired using a structured light scanner (Figure~\ref{fig:DataCollected} (G)). These two point clouds later serve as the target and source surfaces in the deformable registration.

\subsubsection{Fiducial Point Extraction}

For the fiducial points constraint, we manually identified the surgical suture points on both the specimen mesh and resection bed point cloud (Figure~\ref{fig:DataCollected} (C) and (F)) placed prior to resection. These corresponding suture points serve as fiducial points for both the initial rigid registration and the subsequent deformable registration.

\subsubsection{Contour Line Extraction}
We developed an automated pipeline to extract closed 3D contour lines from point cloud data. Given a set of 3D points representing a surface region of interest, the pipeline extracts a closed contour representing the outer boundary of that region. The following steps describe the extraction process.

We first applied Principal Component Analysis (PCA) to the 3D point cloud $\mathbf{P}_{3D}$ to estimate the dominant surface plane. The two principal components corresponding to the largest eigenvalues span a projection plane, approximating the plane of the surface. We projected $\mathbf{P}_{3D}$ onto this plane to obtain the 2D representation $\mathbf{P}_{2D} = \text{PCA}_2(\mathbf{P}_{3D})$, discarding the component of least variance.

We then performed Delaunay triangulation on $\mathbf{P}_{2D}$ to connect all points into a mesh of triangles. From this triangulation, we identified boundary points as vertices lying on the outer polygonal boundary of the triangulation, and boundary edges as edges incident to only one triangle. The raw Delaunay triangulation often includes spurious triangles connecting distant boundary points, creating artificially long edges that do not represent true surface connectivity. To remove these artifacts, we computed a fixed threshold $\tau$ as the 90th percentile of the initial boundary edge lengths and applied iterative refinement. At each iteration $k$, we identified the current boundary edges $E_{\text{boundary}}^{(k)}$ and removed triangles containing edges exceeding $\tau$:
\begin{equation}
T_{\text{remove}}^{(k)} = \{t \in T^{(k)} \mid \exists e \in t \cap E_{\text{boundary}}^{(k)}: L(e) > \tau\}
\end{equation}
The triangulation is updated as $T^{(k+1)} = T^{(k)} \setminus T_{\text{remove}}^{(k)}$, and the process repeats until $T_{\text{remove}}^{(k)} = \emptyset$ or 50 iterations are reached. This progressive removal from the periphery inward ensures a clean boundary by iteratively eliminating outlying triangles and re-evaluating newly exposed edges against the same threshold.

From the refined triangulation, we extracted boundary edges and traced the longest connected component using depth-first graph traversal to form an ordered polyline. We restored 3D coordinates by mapping each 2D boundary point to its nearest neighbor in the original point cloud via k-d tree search: $\mathbf{B}_{3D}[i] = \mathbf{P}_{3D}[\operatorname*{argmin}_k \|\mathbf{b}_i - \mathbf{P}_{2D}[k]\|_2]$. Finally, we smoothed the contour using Savitzky-Golay filtering (window=7, polynomial order=2) in wrap mode and closed the polyline by appending the first point. 

We applied this pipeline to two inputs in this study: manually selected vertices from the posterior surface of the resected specimen mesh (Figure~\ref{fig:DataCollected} (G)) and the resection bed point cloud reconstructed from the stereo camera (Figure~\ref{fig:DataCollected} (C)). This process yields two smooth, closed 3D contours for the specimen boundary (Figure~\ref{fig:DataCollected} (H)) and the resection bed boundary (Figure~\ref{fig:DataCollected} (D)), respectively, which are subsequently used to establish correspondence constraints in the deformable registration framework.

\subsection{Deformation Correction}
To correct the intraoperative deformation on the specimen, we use a sparse data registration algorithm proposed by Ringel et al.~\cite{RingelMorganandHeiselman2023RegularizedBreast} to match the 3D scans of the resected specimen to the resection bed.

Prior to deformable registration, we perform an initial rigid alignment using fiducial points identified on both the specimen (green points in Figure~\ref{fig:DataCollected} (F)) and resection bed (red points in Figure~\ref{fig:DataCollected} (C)) to establish approximate correspondence.

The deformable registration algorithm consists of two phases: the pre-computation phase and the reconstruction phase. In the pre-computation phase, we construct a displacement basis representing the space of possible deformations on the specimen surface using regularized Kelvinlet displacement solutions. These functions are analytical solutions to the Navier-Cauchy equations for linear elasticity in an infinite 3D continuum~\cite{de2017regularized}, and can be scaled and rotated to model point-source tissue deformations of varying magnitudes and directions. We distribute 45 control points on the 3D specimen surface with a radial scale parameter of $\epsilon = 0.01$ m. Regarding material properties, the displacement and strain solutions are independent of Young's modulus under pure displacement boundary conditions, and any difference in tissue stiffness between specimens is compensated by adjusting the strain energy weight $w_E$ at registration time~\cite{heiselman2020intraoperative}.

In the reconstruction phase, we use the linearized iterative boundary reconstruction (LIBR) method~\cite{heiselman2020intraoperative} to calculate a displacement field that registers the 3D specimen mesh to the resection bed point cloud. The deformation field is represented as a linear combination of the regularized Kelvinlet basis vectors, $\tilde{\mathbf{u}} = \mathbf{J}_u^{RK}\boldsymbol{\alpha}$, where $\boldsymbol{\alpha} \in \mathbb{R}^{3k}$ are the combination weights. To account for rigid motion and volumetric shrinkage following resection, we augment the parameter vector to $\boldsymbol{\beta} = [\boldsymbol{\alpha}, \boldsymbol{\tau}, \boldsymbol{\theta}, s]$, incorporating translation $\boldsymbol{\tau}$, rotation $\boldsymbol{\theta}$, and isotropic scale $s$. We find the optimal $\boldsymbol{\beta}$ via Levenberg-Marquardt optimization terminating when $|\Delta\Omega(\boldsymbol{\beta})| < 10^{-12}$, minimizing the objective function:
\begin{equation}
\label{eq:objective}
\Omega(\boldsymbol{\beta}) = \frac{w_{\text{surface}}}{N_{\text{surface}}} \sum_{i=1}^{N_{\text{surface}}} e_i^{\text{surface},2} + \frac{w_{\text{fiducial}}}{N_{\text{fiducial}}} \sum_{k=1}^{N_{\text{fiducial}}} e_k^{\text{fiducial},2} + \frac{w_{\text{contour}}}{N_{\text{contour}}} \sum_{j=1}^{N_{\text{contour}}} e_j^{\text{contour},2} + w_E e_E^2
\end{equation}
The first three terms quantify alignment errors for the surface point clouds, fiducial points, and boundary contour lines, respectively. For surface point clouds, $e_i^{\text{surface}}$ is the normal-projected distance error for each surface point $i$, and $N_{\text{surface}}$ is the number of surface points sampled from the resection bed point cloud. For fiducial landmarks, $e_k^{\text{fiducial}}$ is the Euclidean distance error for each fiducial point $k$, and $N_{\text{fiducial}}$ is the number of fiducial correspondences (typically four per specimen). For boundary contour lines, $e_j^{\text{contour}}$ is the perpendicular distance error for each contour point $j$, and $N_{\text{contour}}$ is the number of contour points. The fourth term penalizes biomechanically implausible deformations, where $e_E$ is the strain energy of the deformation and $w_E$ is the strain energy weighting factor.

\section{Experimental Setup}
\subsection{Parameter Search for Optimization Weights}

The objective function (Equation~\ref{eq:objective}) balances multiple data-driven error terms and strain energy regularization through weights $w_{\text{surface}}$, $w_{\text{fiducial}}$, $w_{\text{contour}}$, and $w_E$. The choice of these weights determines the relative importance of each term on the registration outcome. To characterize how registration accuracy varies with these weights across tissue types, we performed a systematic two-stage parameter search.

We evaluated registration accuracy using Target Registration Error (TRE) with a leave-one-out protocol. The TRE quantifies the Euclidean distance between an excluded target fiducial point on the specimen after deformation correction and its corresponding suture point on the resection bed. For each specimen, we iteratively excluded one fiducial point as the target and performed registration using the remaining fiducial points. The TRE for each specimen is reported as the mean and standard deviation across all excluded target points.

\subsubsection{Stage 1: Strain Energy Weight Selection}

Since the objective function weights are relative rather than absolute, fixing one term reduces the search dimensionality without loss of generality. We fixed $w_{\text{surface}} = 1$ m$^{-2}$ as a reference scale. We constrained $w_{\text{contour}} = w_{\text{fiducial}}$ (ranging from 1 to 10 m$^{-2}$) as a neutral baseline in Stage 1, deferring the exploration of their relative balance to Stage 2. This is motivated by the difference in correspondence type between the two terms: fiducial points are predefined exact correspondences placed prior to resection, while contour points rely on nearest-neighbor matching and carry different uncertainty characteristics. This reduces the Stage 1 search to a two-dimensional space over $w_E$ and $w_{\text{contour}} = w_{\text{fiducial}}$. We varied $w_E$ over a logarithmic range from $10^{-13}$ to $10^{-5}$ Pa$^{-2}$ and computed TRE for each configuration. The value $w_E^*$ was selected as the weight achieving the lowest mean TRE across all specimens: insufficient regularization permits physically implausible deformations, while excessive regularization prevents the algorithm from fitting the observed data.

\subsubsection{Stage 2: Feature Weight Balance}

With $w_{\text{surface}} = 1$ m$^{-2}$ and $w_E = w_E^*$ fixed, we performed a two-dimensional grid search over $w_{\text{contour}}$ (1 to 10 m$^{-2}$) and $w_{\text{fiducial}}$ (1 to 10 m$^{-2}$) to characterize how the relative weighting of the contour and fiducial terms affects registration accuracy. For each parameter combination, we evaluated registration and computed TRE. Results are reported as heat maps with $w_{\text{fiducial}}$ on the x-axis and $w_{\text{contour}}$ on the y-axis for all specimens combined and separately for each tissue type.

This two-stage approach separately addresses regularization strength (Stage 1) and the relative importance of geometric feature terms (Stage 2), providing systematic exploration of the parameter space defined by the objective function.

\subsection{System Setups \& Data Collection}
\subsubsection{System Hardware}
\begin{figure}[h!]
  \centering
  \includegraphics[width=0.9\textwidth]{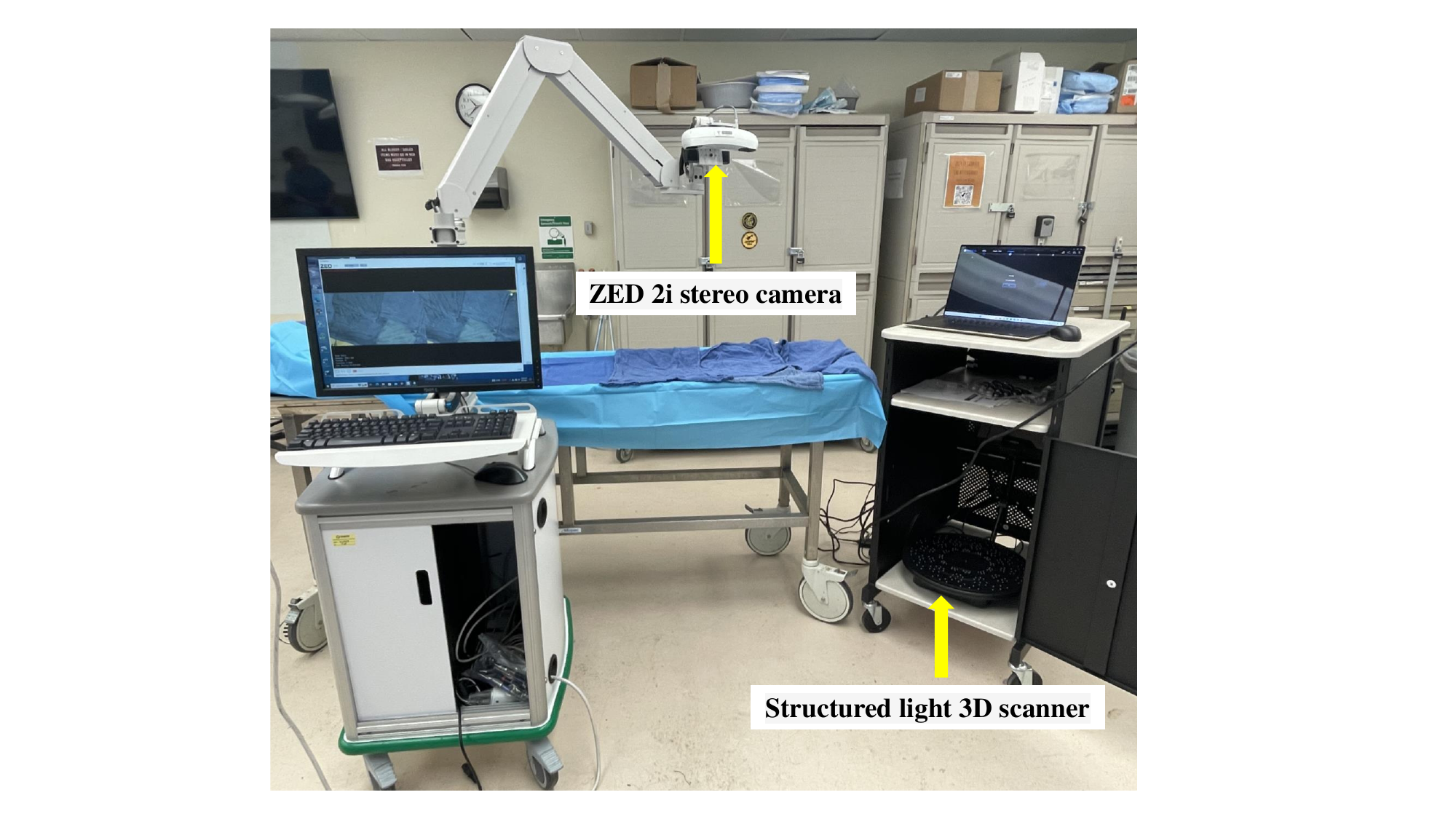}
  \caption{Experimental setup with the ZED camera positioned on the left and the 3D scanner on the right}
  \label{fig:setups}
\end{figure}

Figure~\ref{fig:setups} shows the system hardware setup. The proposed surgical guidance method employs a structured light 3D scanner (EinScan SP, Shining 3D, Hangzhou, China) to create the mesh of the resected specimen following the prior protocol~\cite {perez2023ex} and an RGBD camera, the ZED 2i stereo camera (Stereolabs Inc., San Francisco, CA, USA), to capture the point cloud of the resection bed. The RGBD camera, fixed on an extendable arm and secured to a portable cart, enables data collection from the overhead position in surgical environments~\cite{ringel2024image}. We extract the resection bed point cloud by manually segmenting the region of interest from a single frame of video data captured by the RGBD camera.

\subsubsection{Data Collection}
\label{sec:data_collection}

We conducted a series of experiments using nine head and neck specimens from fresh-frozen human cadaver heads, comprising three specimens each of skin, buccal, and tongue. We collected these specimens in two previous studies~\cite{Yang2025Nonrigid, yang2025augmented} using the same experimental protocol. The assumption of our study is that the resected specimens from these cadavers experience mucosal shrinkage similar to that observed in frozen section specimens from living patients. This study investigated cadavers with approval from the Vanderbilt University Medical Center’s Center for Experiential Learning and Assessment. 

A fellowship-trained head and neck cancer surgeon, who performs approximately 50 locally advanced head and neck surgical resections annually, conducted the resections during the cadaver studies. Prior to resection, the surgeon used an ink pen to mark the surgical plan (Figure~\ref{fig:DataCollected} (A)).
During the procedure, they sutured 4 pairs of corresponding points on the resection bed and the specimen's edge (visible in Figure~\ref{fig:DataCollected} (B) and (E)), which served as fiducial points. After the electrocautery resections, we used the RGBD camera to capture the geometric data of the resection bed. Following the ex vivo 3D scanning protocol~\cite{perez2023ex}, we obtained the 3D mesh of the resected specimen (Figure~\ref{fig:DataCollected} (F)).

\section{Results}
\label{sec:results}

\subsection{Impact of contour constraint}

Table~\ref{tab:tre_comparison} compares registration accuracy across three methods: rigid registration using only fiducial correspondences, deformable registration without contour constraint, and deformable registration with contour constraint. To evaluate the contribution of boundary contour information to registration performance, we performed the deformable registration with and without the contour constraint using an equal-weighted baseline configuration: $w_E = 10^{-9}$ Pa$^{-2}$, $w_{\text{surface}} = w_{\text{fiducial}} = 1$ m$^{-2}$. Deformable registration with the contour constraint used the same configuration with $w_{\text{contour}} = 1$.

Rigid registration achieved an overall mean TRE of 11.11 mm. Deformable registration without contour constraint reduced overall mean TRE to 8.20 mm (26.19\% reduction from rigid). Deformable registration with contour constraint, using $w_E = 10^{-9}$ Pa$^{-2}$, $w_{\text{surface}} = w_{\text{fiducial}} = w_{\text{contour}} = 1$ m$^{-2}$, further reduced overall mean TRE to 6.13 mm (25.2\% reduction from without contour; 44.8\% reduction from rigid). 

The reduction in TRE varied across tissue types, with tongue specimens posing the greatest clinical challenge in margin relocalization and demonstrating the largest improvement (37.6\%). The contour constraint reduced or maintained TRE across all specimens.

The improved performance with contour constraint indicates that boundary contour information contributes to registration accuracy beyond what surface and fiducial terms alone provide. Since the magnitude of this contribution depends on contour weighting, we performed a systematic parameter search to characterize how contour, fiducial, surface, and strain energy weights interact, as reported in the following section.

\begin{table}[ht]
\caption{Target Registration Error (TRE) of all specimens. Comparison of rigid registration, deformable registration without contour constraint, and deformable registration with contour constraint. The deformable registration without contour configuration uses parameters $w_E = 10^{-9}$ Pa$^{-2}$, $w_{\text{surface}} = 1$ m$^{-2}$, $w_{\text{fiducial}} = 1$ m$^{-2}$; the deformable registration with contour configuration additionally includes $w_{\text{contour}} = 1$ m$^{-2}$. Individual specimen values reported as mean $\pm$ standard deviation in mm across four leave-one-out fiducial points. Average and overall values reported as within group pooled mean $\pm$ pooled standard deviation in mm. Bold values indicate the better performance.} 
\label{tab:tre_comparison}
\begin{center}       
\begin{tabular}{|l|c|c|c|} 
\hline
\rowcolor[gray]{0.85}
\rule[-1ex]{0pt}{3.5ex}  Specimen & Rigid Registration & Without Contour Constraint & With Contour Constraint  \\
\hline\hline
\rule[-1ex]{0pt}{3.5ex}  Skin-1 & $11.25\pm1.48$ & $7.50\pm2.31$ & $\mathbf{6.09\pm2.85}$  \\
\hline
\rule[-1ex]{0pt}{3.5ex}  Skin-2 & $9.96\pm1.09$ & $6.31\pm1.19$ & $\mathbf{6.29\pm1.92}$   \\
\hline
\rule[-1ex]{0pt}{3.5ex}  Skin-3 & $10.79\pm3.85$ & $5.82\pm2.68$ & $\mathbf{4.66\pm1.49}$   \\
\hline
\rowcolor[gray]{0.85}
\rule[-1ex]{0pt}{3.5ex}  Skin Average & $10.67\pm2.46$ & $6.54\pm2.16$ & $\mathbf{5.68\pm2.16}$   \\
\hline\hline
\rule[-1ex]{0pt}{3.5ex}  Buccal-1 & $6.97\pm3.10$ & $6.33\pm2.25$ & $\mathbf{4.68\pm1.62}$  \\
\hline
\rule[-1ex]{0pt}{3.5ex}  Buccal-2 & $11.00\pm1.68
$ & $6.89\pm2.76$ & $\mathbf{5.86\pm0.76}$  \\
\hline
\rule[-1ex]{0pt}{3.5ex}  Buccal-3 & $10.47\pm5.99
$ & $10.19\pm4.27$ & $\mathbf{8.40\pm3.65}$  \\
\hline
\rowcolor[gray]{0.85}
\rule[-1ex]{0pt}{3.5ex}  Buccal Average & $9.48\pm4.01$ &  $7.80\pm3.21$ & $\mathbf{6.31\pm2.35}$  \\
\hline\hline
\rule[-1ex]{0pt}{3.5ex}  Tongue-1 & $15.42\pm2.37$ & $16.48\pm3.85$ & $\mathbf{8.93\pm0.66}$  \\
\hline
\rule[-1ex]{0pt}{3.5ex}  Tongue-2 & $9.18\pm4.62$ & $5.28\pm1.11$ & $\mathbf{4.84\pm2.22}$  \\
\hline
\rule[-1ex]{0pt}{3.5ex}  Tongue-3 & $14.94\pm7.44$ & $9.05\pm2.02$ & $\mathbf{5.45\pm3.76}$ \\
\hline
\rowcolor[gray]{0.85}
\rule[-1ex]{0pt}{3.5ex}  Tongue Average & $13.18\pm5.24$ & $10.27\pm2.59$ & $\mathbf{6.41\pm2.55}$ \\
\hline\hline
\rule[-1ex]{0pt}{3.5ex}  Overall & $11.11\pm4.07$ & $8.20\pm2.68$ & $\mathbf{6.13\pm2.36}$ \\
\hline 
\end{tabular}
\end{center}
\end{table}

\subsection{Parameter Search}
\subsubsection{Stage 1: Strain Energy Weight Selection}
\begin{figure}[htbp]
  \centering
  \includegraphics[width=1\textwidth]{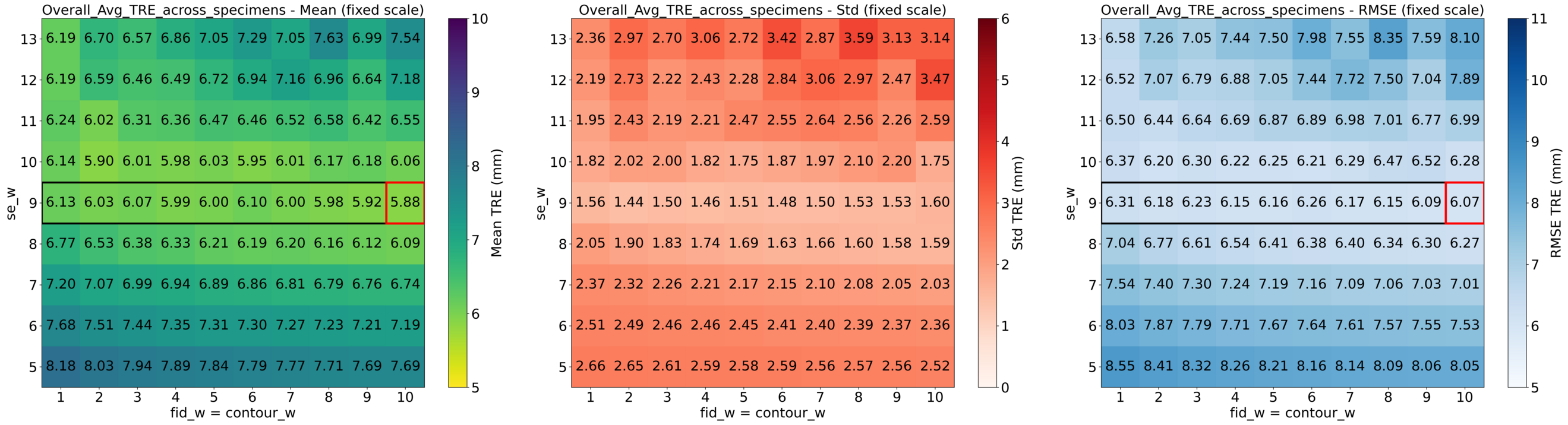}
  \caption{Stage 1 Strain Energy Weight Selection results for overall performance across all specimens. Heat maps show mean target registration error (TRE), standard deviation, and root mean square error (RMSE) as functions of strain energy weight ($w_E$, y-axis, from $10^{-5}$ to $10^{-13}$ Pa$^{-2}$) and equal feature weights ($w_{\text{fiducial}} = w_{\text{contour}}$, x-axis, 1-10 m$^{-2}$). The black box marks the row with the lowest sum of mean/RMSE TRE across all feature weight values. The red box marks the single parameter combination achieving the lowest mean/RMSE TRE in that heat map. Figure~\ref{fig:overall_1} to Figure~\ref{fig:tongue_2} use consistent fixed color scales: mean TRE [5.0-10.0 mm], std [0.0-6.0 mm], RMSE [5.0-11.0 mm] to enable direct comparison across tissue types and parameter search stages.}
  \label{fig:overall_1}
\end{figure}

\begin{figure}[htbp]
  \centering
  \includegraphics[width=\textwidth]{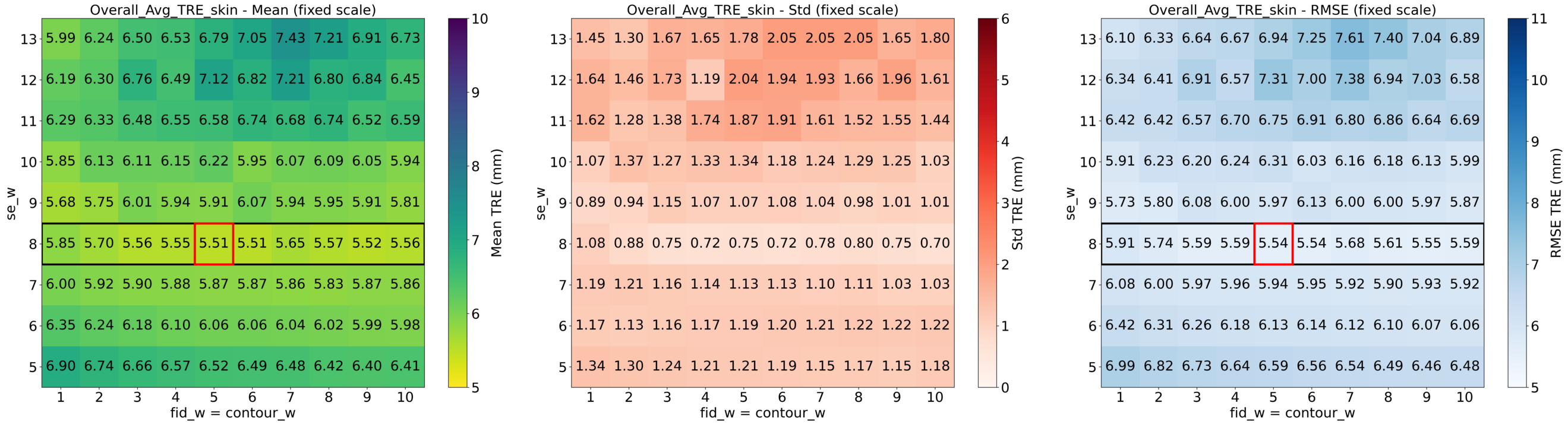}
  \caption{Stage 1 Strain Energy Weight Selection results for skin specimens.The black box marks the row with the lowest sum of mean/RMSE TRE across all feature weight values. The red box marks the single parameter combination achieving the lowest mean/RMSE TRE in that heat map.}
  \label{fig:skin_1}
\end{figure}

\begin{figure}[htbp]
  \centering
  \includegraphics[width=\textwidth]{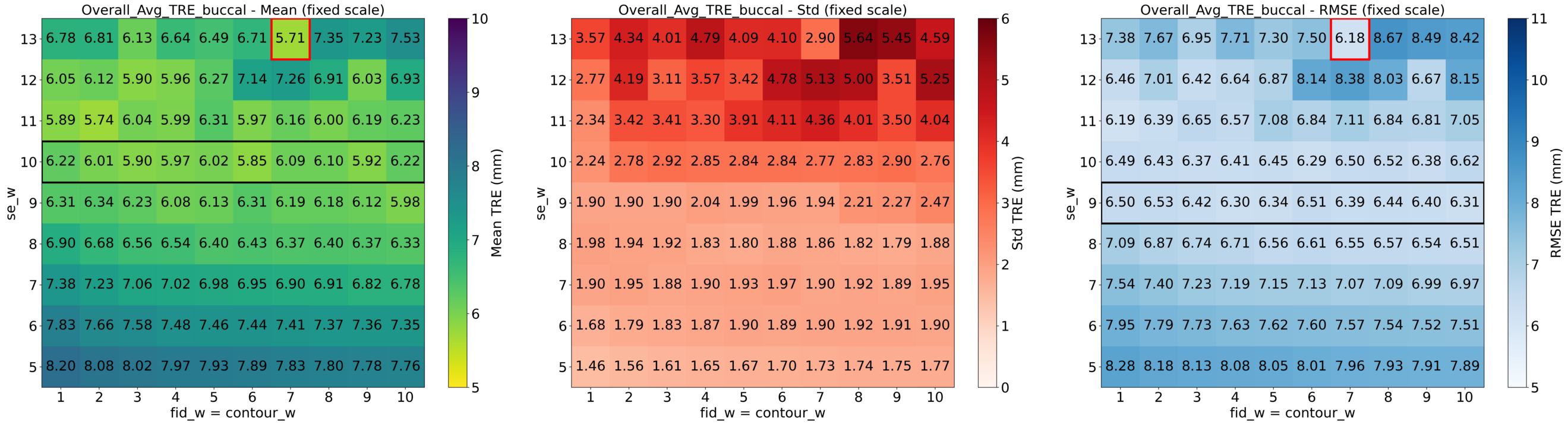}
  \caption{Stage 1 Strain Energy Weight Selection results for buccal specimens. The black box marks the row with the lowest sum of mean/RMSE TRE across all feature weight values. The red box marks the single parameter combination achieving the lowest mean/RMSE TRE in that heat map.}
  \label{fig:buccal_1}
\end{figure}

\begin{figure}[htbp]
  \centering
  \includegraphics[width=\textwidth]{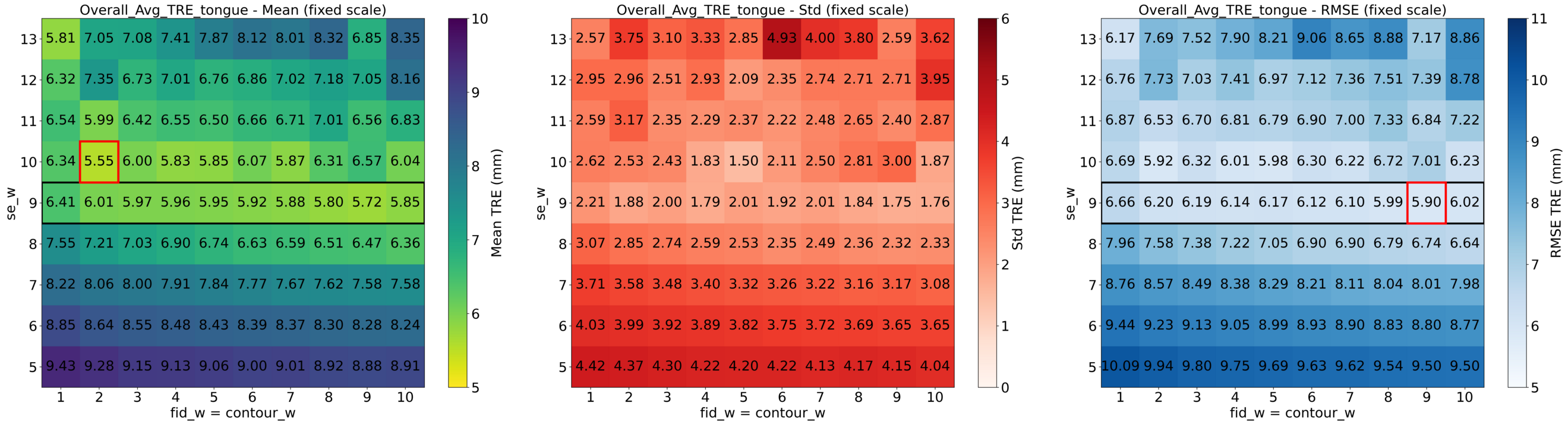}
  \caption{Stage 1 Strain Energy Weight Selection results for tongue specimens. The black box marks the row with the lowest sum of mean/RMSE TRE across all feature weight values. The red box marks the single parameter combination achieving the lowest mean/RMSE TRE in that heat map.}
  \label{fig:tongue_1}
\end{figure}

In this stage, we evaluated registration performance across the parameter space defined by strain energy weight $w_E$ ($10^{-13}$ to $10^{-5}$ Pa$^{-2}$) and equal weights $w_{\text{fiducial}} = w_{\text{contour}}$ (1 to 10 m$^{-2}$), with $w_{\text{surface}}$ fixed at 1 m$^{-2}$. Figure~\ref{fig:overall_1} shows the TRE heat maps aggregated across all 9 specimens. Figure~\ref{fig:skin_1} to Figure~\ref{fig:tongue_1} show the corresponding results of skin, buccal, and tongue (3 specimens each), respectively. Each cell reports the mean TRE, standard deviation, or RMSE TRE computed across specimens for that parameter combination, with rows corresponding to a fixed $w_E$ value. The black box marks the row with the lowest sum of TRE across all feature weight values. The red box marks the single parameter combination achieving the lowest TRE in that heat map.

For the overall specimen performance (Figure~\ref{fig:overall_1}), mean and RMSE TRE both reach their minimum at $w_E = 10^{-9}$ Pa$^{-2}$, $w_{\text{fiducial}} = w_{\text{contour}} = 10$ m$^{-2}$, with $w_E = 10^{-9}$ Pa$^{-2}$ also achieving the minimum row sum. 

Individual tissue types demonstrates varying sensitivity to regularization strength. For skin specimens, the minimum row sum occurrs at $w_E = 10^{-8}$ Pa$^{-2}$ in both the mean and RMSE heat maps. Skin specimens also shows stable performance across the $w_E = 10^{-9}$ to $10^{-7}$ Pa$^{-2}$ range, with all mean TRE below 6.1 mm. For buccal specimens, the minimum row sum appears at $w_E = 10^{-10}$ Pa$^{-2}$ in the mean TRE heat map but shifts to $w_E = 10^{-9}$ Pa$^{-2}$ in the RMSE heat map. Tongue specimens achieves the minimum row sum at $w_E = 10^{-9}$ Pa$^{-2}$ in both the mean and RMSE heat maps.

\subsubsection{Stage 2: Feature Weight Balance}

\begin{figure}[htbp]
  \centering
  \includegraphics[width=\textwidth]{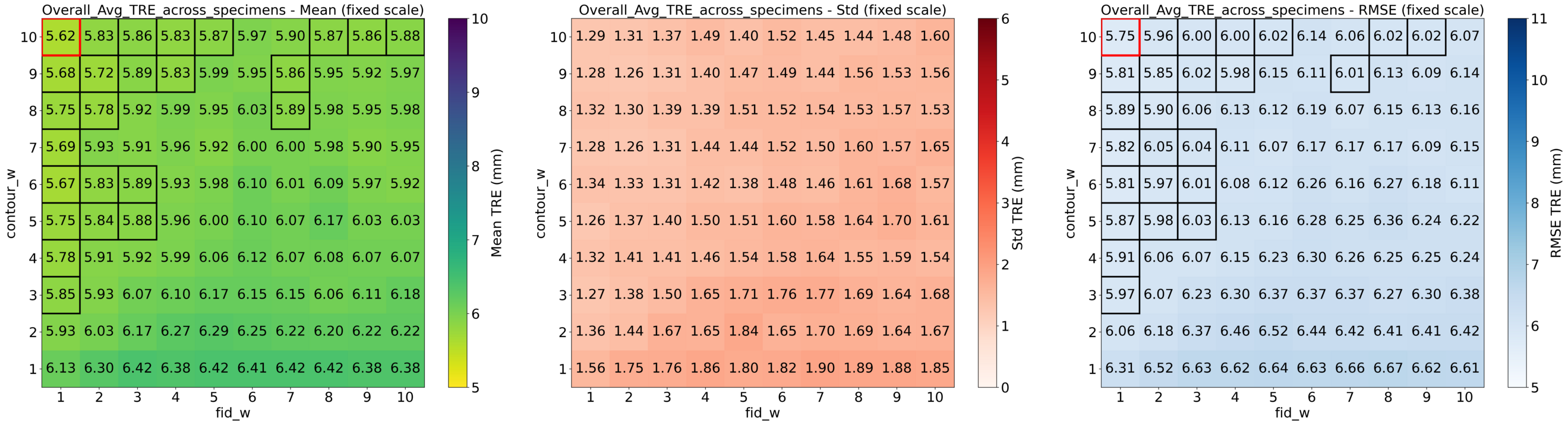}
  \caption{Stage 2 Feature Weight Balance results for overall performance across all specimens. Heat maps show mean target registration error (TRE), standard deviation, and root mean square error (RMSE) as functions of contour weight ($w_{\text{contour}}$, y-axis, 1-10 m$^{-2}$) and fiducial weight ($w_{\text{fiducial}}$, x-axis, 1-5 m$^{-2}$) with strain energy weight fixed at $w_E = 10^{-9}$ Pa$^{-2}$. The red box in the heat maps marks the lowest mean/RMSE TRE, while the black box marks all parameter combinations where mean/RMSE TRE did not exceed 5\% above the minimum.}
  \label{fig:overall_2}
\end{figure}

\begin{figure}[htbp]
  \centering
  \includegraphics[width=\textwidth]{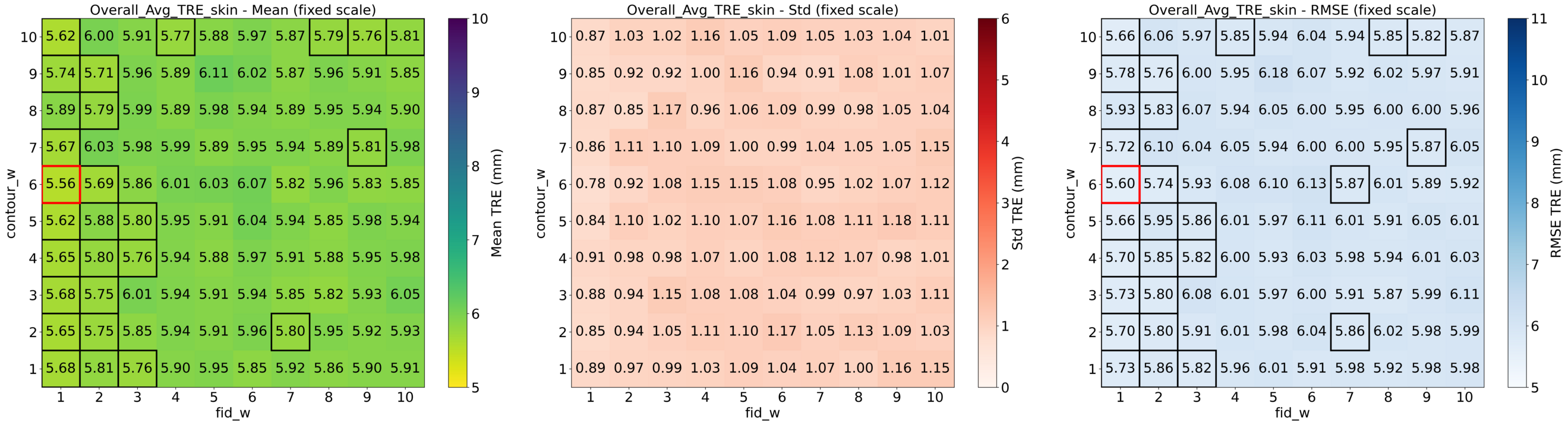}
  \caption{Stage 2 Feature Weight Balance results for skin specimens. The red box in the heat maps marks the lowest mean/RMSE TRE, while the black box marks all parameter combinations where mean/RMSE TRE did not exceed 5\% above the minimum.}
  \label{fig:skin_2}
\end{figure}

\begin{figure}[htbp]
  \centering
  \includegraphics[width=\textwidth]{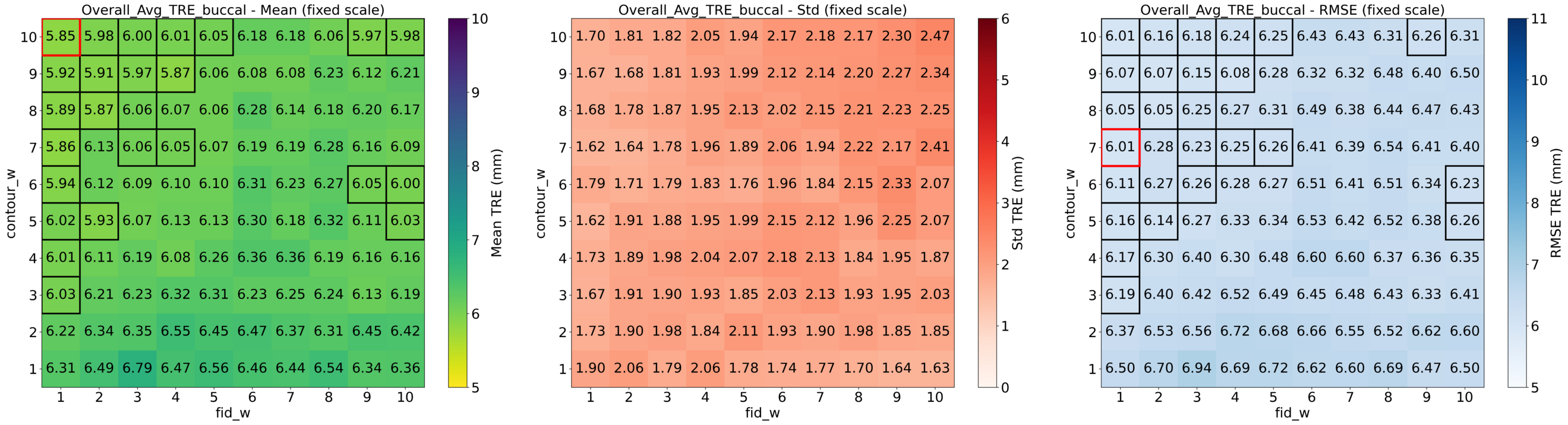}
  \caption{Stage 2 Feature Weight Balance results for buccal specimens. The red box in the heat maps marks the lowest mean/RMSE TRE, while the black box marks all parameter combinations where mean/RMSE TRE did not exceed 5\% above the minimum.}
  \label{fig:buccal_2}
\end{figure}

\begin{figure}[htbp]
  \centering
  \includegraphics[width=\textwidth]{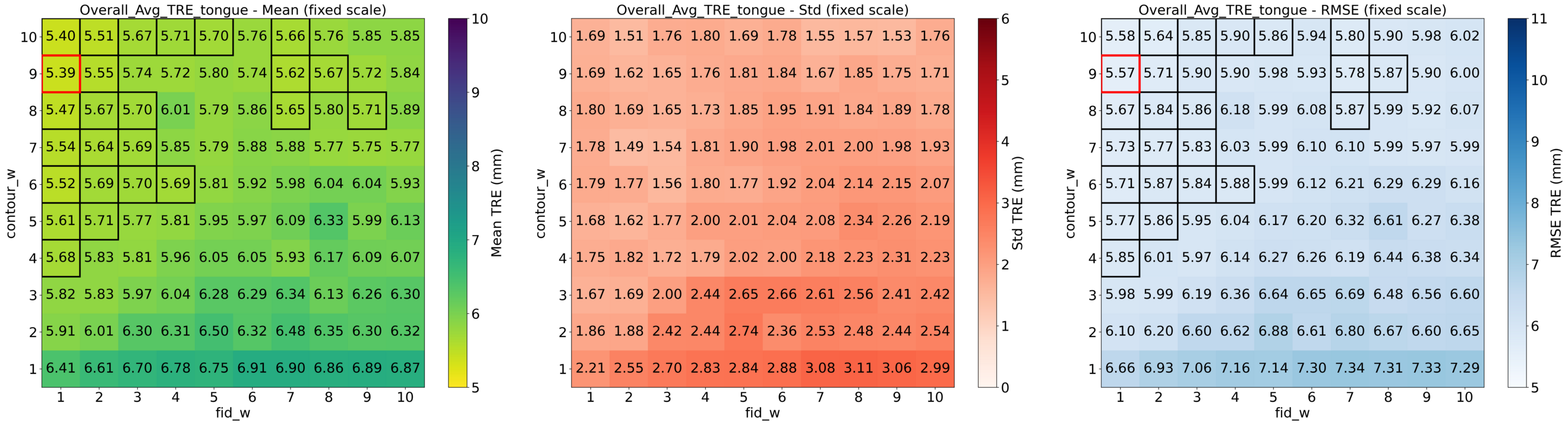}
  \caption{Stage 2 Feature Weight Balance results for tongue specimens. The red box in the heat maps marks the lowest mean/RMSE TRE, while the black box marks all parameter combinations where mean/RMSE TRE did not exceed 5\% above the minimum.}
  \label{fig:tongue_2}
\end{figure}

Figure~\ref{fig:overall_2} to Figure~\ref{fig:tongue_2} shows the resulting heat maps for a grid search over $w_{\text{contour}}$ (1 to 10 m$^{-2}$) and $w_{\text{fiducial}}$ (1 to 5 m$^{-2}$) with fixed $w_E^* = 10^{-9}$ Pa$^{-2}$ and $w_{\text{surface}} = 1.0$ m$^{-2}$. The red box in the heat maps marks the lowest mean/RMSE TRE, while the black box marks all parameter combinations where mean/RMSE TRE does not exceed 5\% above the minimum.

Across all specimens(Figure~\ref{fig:overall_2}), the minimum mean TRE occurrs at $w_{\text{contour}} =$ 10 m$^{-2}$, $w_{\text{fiducial}} = $ 1 m$^{-2}$. The black boxes extend along the high-$w_{\text{contour}}$ edge of the parameter space, with performance decreasing as $w_{\text{fiducial}}$ increased beyond 3 m$^{-2}$.

For skin specimens, the heat maps exhibits the most distributed region within 5\% of the minimum TRE value, with a focus on $w_{\text{fiducial}} < 3 m^{-2}$. For buccal and tongue specimens, the region within 5\% to optimal value is concentrated at high $w_{\text{contour}}$ and low $w_{\text{fiducial}}$, consistent with the overall pattern.

\subsection{Effect of Weights on Registration Accuracy}
Based on the parameter search results, we selected the configuration achieving the lowest overall mean and RMSE TRE: $w_E = 10^{-9}$ Pa$^{-2}$, $w_{\text{surface}} = 1$ m$^{-2}$, $w_{\text{fiducial}} = 1$ m$^{-2}$, $w_{\text{contour}} = 10$ m$^{-2}$. 

Table~\ref{tab:tre_comparison_after} reports TRE for all specimens under three configurations: without contour constraint, with $w_{\text{contour}} = 1$ m$^{-2}$, and with $w_{\text{contour}} = 10$ m$^{-2}$. Increasing $w_{\text{contour}}$ from 1 to 10 m$^{-2}$ reduces overall mean TRE from $6.13 \pm 2.36$ mm to $5.62 \pm 2.28$ mm, with improvements across all tissue type averages. The largest reduction from $w_{\text{contour}} = 1$ to $w_{\text{contour}} = 10$ m$^{-2}$ occurs in tongue specimens, from $6.41 \pm 2.55$ mm to $5.40 \pm 2.30$ mm. Compared to registration without contour constraint, this represents a 37.6\% reduction at $w_{\text{contour}} = 1$ and a 47.4\% reduction at $w_{\text{contour}} = 10$.

At the individual specimen level, two skin specimens (Skin-2 and Skin-3) shows lower TRE at $w_{\text{contour}} = 1$ m$^{-2}$ than at $w_{\text{contour}} = 10$ m$^{-2}$. 

\begin{figure}[htbp]
  \centering
  \includegraphics[width=\textwidth]{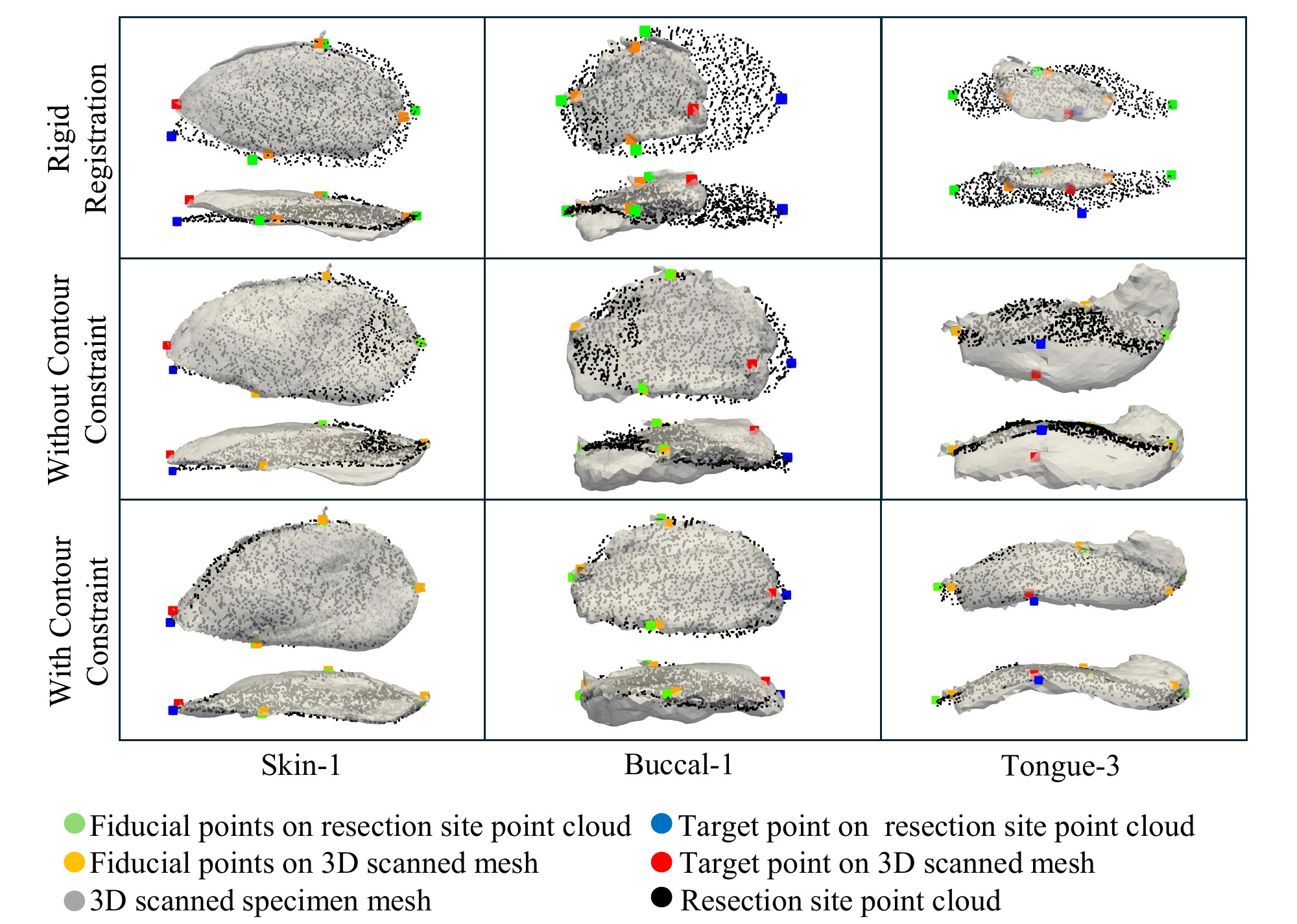}
  \caption{Qualitative registration results for a representative leave-one-out trial of Skin-1, Buccal-1, and Tongue-3 (left to right). The first row shows rigid registration between the 3D specimen mesh and the resection bed point cloud. The second row shows deformable registration without the contour constraint ($w_E = 10^{-9}$ Pa$^{-2}$, $w_{\text{surface}} = w_{\text{fiducial}} = 1$ m$^{-2}$, $w_{\text{contour}} = 0$). The third row shows the proposed deformable registration with the contour constraint ($w_E = 10^{-9}$ Pa$^{-2}$, $w_{\text{surface}} = w_{\text{fiducial}} = 1$ m$^{-2}$, $w_{\text{contour}} = 10$ m$^{-2}$).}
  \label{fig:contour_compare}
\end{figure}

Figure~\ref{fig:contour_compare} shows qualitative registration results for a representative leave-one-out trial of Skin-1, Buccal-1, and Tongue-3, comparing rigid registration, deformable registration without the contour constraint, and the proposed deformable registration with $w_{\text{contour}} = 10$ m$^{-2}$. After rigid registration, Buccal-1 and Tongue-3 display larger misalignment between the specimen mesh and the resection bed point cloud, indicating that these specimens underwent larger post-resection deformation compared to skin specimens. The second row of Figure~\ref{fig:contour_compare} shows that deformable registration without the contour constraint reduces this misalignment, but lateral offset between the specimen mesh boundary and the resection bed point cloud remains large, especially in Tongue-3. With $w_{\text{contour}} = 10$ m$^{-2}$, the specimen mesh boundary aligned more closely with the resection bed point cloud boundary across all three specimens. These qualitative observations are consistent with the TRE reductions reported in Table~\ref{tab:tre_comparison_after}.

\begin{table}[ht]
\caption{Target Registration Error (TRE) of all specimens, comparing registration performance without contour constraint and with two contour weight settings. The without-contour configuration uses parameters $w_E = 10^{-9}$ Pa$^{-2}$, $w_{\text{surface}} = 1$ m$^{-2}$, $w_{\text{fiducial}} = 1$ m$^{-2}$; the with-contour configurations additionally include $w_{\text{contour}} = 1$ m$^{-2}$ and $w_{\text{contour}} = 10$ m$^{-2}$ respectively. Individual specimen values reported as mean $\pm$ standard deviation in mm across four leave-one-out fiducial points. Average and overall values reported as pooled mean $\pm$ pooled standard deviation in mm. Bold values indicate the best-performing configuration per specimen.}
\label{tab:tre_comparison_after}
\begin{center}       
\begin{tabular}{|l|c|c|c|} 
\hline
\rowcolor[gray]{0.85}
\rule[-1ex]{0pt}{3.5ex}  Specimen & Without Contour & $w_{\text{contour}} = 1$ & $w_{\text{contour}} = 10$  \\
\hline\hline
\rule[-1ex]{0pt}{3.5ex}  Skin-1 & $7.50\pm2.31$ & $6.09\pm2.85$ & $\mathbf{5.73\pm2.62}$  \\
\hline
\rule[-1ex]{0pt}{3.5ex}  Skin-2 & $6.31\pm1.19$ & $\mathbf{6.29\pm1.92}$ & $6.43\pm1.75$  \\
\hline
\rule[-1ex]{0pt}{3.5ex}  Skin-3 & $5.82\pm2.68$ & $\mathbf{4.66\pm1.49}$ & $4.70\pm1.30$  \\
\hline
\rowcolor[gray]{0.85}
\rule[-1ex]{0pt}{3.5ex}  Skin Average & $6.54\pm2.16$ & $5.68\pm2.16$ & $\mathbf{5.62\pm1.97}$  \\
\hline\hline
\rule[-1ex]{0pt}{3.5ex}  Buccal-1 & $6.33\pm2.25$ & $4.68\pm1.62$ & $\mathbf{4.25\pm1.14}$  \\
\hline
\rule[-1ex]{0pt}{3.5ex}  Buccal-2 & $6.89\pm2.76$ & $5.86\pm0.76$ & $\mathbf{5.64\pm0.96}$  \\
\hline
\rule[-1ex]{0pt}{3.5ex}  Buccal-3 & $10.19\pm4.27$ & $8.40\pm3.65$ & $\mathbf{7.64\pm4.15}$  \\
\hline
\rowcolor[gray]{0.85}
\rule[-1ex]{0pt}{3.5ex}  Buccal Average & $7.80\pm3.21$ & $6.31\pm2.35$ & $\mathbf{5.84\pm2.55}$  \\
\hline\hline
\rule[-1ex]{0pt}{3.5ex}  Tongue-1 & $16.48\pm3.85$ & $8.93\pm0.66$ & $\mathbf{7.32\pm0.84}$  \\
\hline
\rule[-1ex]{0pt}{3.5ex}  Tongue-2 & $5.28\pm1.11$ & $4.84\pm2.22$ & $\mathbf{4.12\pm2.28}$  \\
\hline
\rule[-1ex]{0pt}{3.5ex}  Tongue-3 & $9.05\pm2.02$ & $5.45\pm3.76$ & $\mathbf{4.77\pm3.15}$  \\
\hline
\rowcolor[gray]{0.85}
\rule[-1ex]{0pt}{3.5ex}  Tongue Average & $10.27\pm2.59$ & $6.41\pm2.55$ & $\mathbf{5.40\pm2.30}$  \\
\hline\hline
\rule[-1ex]{0pt}{3.5ex}  Overall & $8.20\pm2.68$ & $6.13\pm2.36$ & $\mathbf{5.62\pm2.28}$  \\
\hline 
\end{tabular}
\end{center}
\end{table}

\section{Discussion}
\label{sec:discussion}

\subsection{Impact of contour constraint and Weight Selection}

The reduction in overall mean TRE from rigid registration to deformable registration, both with and without contour constraint (Table~\ref{tab:tre_comparison}), highlights the importance of modeling post-resection tissue deformation in head and neck specimens. While deformable registration without contour constraint improved TRE relative to rigid registration, incorporating boundary contour information further reduced mean TRE across all specimens. These findings suggest the effectiveness of incorporating contour constraint in deformable registration. 

Without contour constraint, surface point clouds enforce alignment through normal-direction distances, which constrain local surface curvature but do not penalize lateral boundary displacement. Fiducial points provide only sparse correspondence, and the number of available points is further limited by specimen size, as smaller specimens cannot support many suture placements. As a result of point-to-surface correspondence errors being projected onto the surface normal direction, specimen meshes with thin and high curvature edges may incompletely match the resection bed boundary while still achieving low surface and fiducial error, as observed in Tongue-3 in Figure~\ref{fig:contour_compare}. This violates the physical constraint that the registered specimen should remain within the resection cavity and reduces the accuracy of target point localization near the boundary.

The contour constraint addresses this by directly penalizing distance-to-agreement along the resection margin. The tissue-specific pattern in Table~\ref{tab:tre_comparison} is consistent with this mechanism: tongue specimens, which underwent the largest post-resection deformation, showed the largest TRE reduction (37.6\%), while skin specimens, which deformed less, showed the smallest reduction. This also explains why increasing $w_{\text{contour}}$ from 1 to 10 m$^{-2}$ yielded further improvement in Table~\ref{tab:tre_comparison_after}. At low contour weighting, the contour term is insufficient to correct boundary placement when it conflicts with surface and fiducial terms. The additional reduction was again largest for tongue specimens, consistent with the greater importance of boundary information for tissue with large post-resection deformation.

The two skin specimens (Skin-2 and Skin-3) that performed slightly better at $w_{\text{contour}} = 1$ m$^{-2}$ than at $w_{\text{contour}} = 10$ m$^{-2}$ suggest that a high contour weight is not universally beneficial. Unlike fiducial points, which provide exact point-to-point correspondences, the contour constraint relies on nearest-neighbor matching between contour polylines, introducing approximation error when the true correspondence is ambiguous. For buccal and tongue specimens, which exhibit deformation in both lateral and thickness directions as shown in the first row of Figure~\ref{fig:contour_compare}, the contour constraint provides boundary information that surface and fiducial terms cannot capture, and the approximation error is outweighed by the geometric guidance it provides. For skin specimens, which undergo primarily lateral deformation with limited thickness change, the deformation pattern is already well-captured by surface and fiducial terms. In this case, placing high weight on the approximate contour correspondences over-constrains the registration relative to the exact fiducial correspondences, potentially introducing error rather than reducing it.

\subsection{Parameter Search}

The two-stage parameter search revealed how the relative importance of each objective term varies with regularization strength and tissue type, providing a basis for parameter selection in contour-constrained deformable registration.

In Stage 1, the aggregated optimal strain energy weight $w_E^* = 10^{-9}$ Pa$^{-2}$ reflects a balance between data fidelity and deformation regularity: weaker regularization permitted physically implausible deformations, while stronger regularization over-constrained the solution. The tissue-specific variation in optimal $w_E$ is consistent with differences in post-resection deformation magnitude. Skin specimens favored stronger regularization ($w_E = 10^{-8}$ Pa$^{-2}$), suggesting smaller and more uniform deformation. Buccal and tongue specimens, which are highly deformable, performed best at $w_E = 10^{-10} $ to $10^{-9}$ Pa$^{-2}$ for mean and RMSE TRE, requiring weaker regularization to accommodate larger shape changes.

In Stage 2, the concentration of the competitive region at high $w_{\text{contour}}$ and low $w_{\text{fiducial}}$ across all tissue types indicates that boundary contour information provided the dominant geometric constraint for registration under these experimental conditions. This is consistent with the information content of each term: fiducial points provide only four sparse point correspondences per specimen, while the contour constraint enforces alignment along a continuous boundary curve. Furthermore, fiducial localization errors due to manual suture placement at corresponding points between the specimen and the resection bed and their subsequent digitization is likely higher than the RGBD camera measurement errors of the cavity surface and the margin contour. The broad competitive region observed for skin specimens, which distributed across a wide range of $w_{\text{contour}}$ at low $w_{\text{fiducial}}$, further supports the interpretation that contour constraint are most informative for tissue types with larger and more complex post-resection deformation. For buccal and tongue specimens, the narrower competitive region concentrated at high $w_{\text{contour}}$ suggests that accurate contour weighting is more critical for achieving good registration in these tissue types.

Overall, the parameter search characterizes the sensitivity of the registration algorithm to objective term weighting across different tissue types, rather than identifying a single optimal configuration. Across all specimens, the competitive region in Stage 2 covered a broad range of parameter combinations at high $w_{\text{contour}}$ and low $w_{\text{fiducial}}$, indicating that the algorithm operates reliably over a wide parameter range when tissue types are pooled. This robustness is practically important for clinical deployment, where a single parameter configuration should generalize across specimens without tissue-type-specific tuning. However, when partitioned by tissue type, the optimal parameter regions diverged: skin specimens showed a broad and distributed competitive region, while buccal and tongue specimens showed a narrower region concentrated at high $w_{\text{contour}}$. This tissue-specific sensitivity suggests that the algorithm is well-suited to the head and neck registration problem, where different tissue types present distinct deformation patterns that the objective function can capture through appropriate weighting. Future work should focus on deeper characterization of tissue-related differences in deformation behavior, which could inform tissue-type-specific parameter selection or adaptive weighting strategies that adjust to specimen geometry without requiring manual tuning.

\section{Conclusion}
\label{sec:conclusion}
This work introduced boundary contour constraint into a regularized Kelvinlet-based deformable registration framework for intraoperative head and neck surgical guidance, and performed a systematic two-stage parameter search to characterize how surface alignment, fiducial correspondences, contour constraint, and strain energy regularization interact across tissue types and regularization regimes. With an equal-weighted baseline configuration, incorporating contour constraint reduced the overall mean TRE from $8.20$ mm using deformable registration without contour constraint to $6.13$ mm. The largest reduction in TRE appeared in tongue specimens, which exhibited the greatest post-resection deformation and represent the most clinically challenging cases. Increasing contour weighting to the parameter-search optimal further reduced overall mean TRE to $5.62$ mm. 

The parameter search revealed that the relative importance of each objective term varies with tissue type and deformation magnitude, with contour information contributing most for specimens exhibiting large lateral deformation and fiducial correspondences playing a greater role for specimens with small, uniform deformation. Despite tissue-specific variations in geometry and deformability, the proposed approach operated reliably across a broad range of parameter settings, suggesting that it is well suited to the head and neck registration problem. These findings characterize the sensitivity of the registration algorithm to objective-term weighting across tissue types and provide a foundation for parameter selection in future clinical deployment. 

\section*{Acknowledgements} 
This work was supported in part by the National Institute of Biomedical Imaging and Bioengineering (NIBIB) and the National Cancer Institute (NCI) of the National Institutes of Health (NIH) under Award Numbers R01EB027498, R01EB037685, and K08CA293255, as well as the Vanderbilt University Seeding Success program.

\section*{Code and Data Availability} 
The source code developed for this work will not be publicly released because it relies on proprietary third-party libraries that are not publicly available. The data used in this study are subject to access restrictions.

\section*{Disclosures} 
The authors declare there are no financial interests, commercial affiliations, or other potential conflicts of interest that have influenced the objectivity of this research or the writing of this paper.


\bibliography{report}   

@inproceedings{Yang2025Nonrigid,
  title={Nonrigid Alignment of En Bloc Tissue Specimen to Resection Bed to Enhance Correspondence for Re-Resection Guidance},
  author={Yang, Qingyun and Acar, Ayberk and Ringel, Morgan J and Heiselman, Jon S and Miga, Michael I and Topf, Michael and Wu, Jie Ying},
  booktitle={Medical imaging 2025: image-guided procedures, robotic interventions, and modeling},
  year={2025},
  organization={SPIE}
}

@inproceedings{yang2025augmented,
  title={Augmented Reality-Based Guidance with Deformable Registration in Head and Neck Tumor Resection},
  author={Yang, Qingyun and Li, Fangjie and Xu, Jiayi and Liu, Zixuan and Sridhar, Sindhura and Jin, Whitney and Du, Jennifer and Heiselman, Jon and Miga, Michael and Topf, Michael and others},
  booktitle={International Conference on Medical Image Computing and Computer-Assisted Intervention},
  pages={45--54},
  year={2025},
  organization={Springer}
}

@article{prasad2024more,
  title={More than meets the eye: Augmented reality in surgical oncology},
  author={Prasad, Kavita and Fassler, Carly and Miller, Alexis and Aweeda, Marina and Pruthi, Sumit and Fusco, Joseph C and Daniel, Bruce and Miga, Michael and Wu, Jie Ying and Topf, Michael C},
  journal={Journal of Surgical Oncology},
  volume={130},
  number={3},
  pages={405--418},
  year={2024},
  publisher={Wiley Online Library}
}

@article{tong2024development,
  title={Development of an augmented reality guidance system for head and neck cancer resection},
  author={Tong, Guansen and Xu, Jiayi and Pfister, Michael and Atoum, Jumanh and Prasad, Kavita and Miller, Alexis and Topf, Michael and Wu, Jie Ying},
  journal={Healthcare Technology Letters},
  volume={11},
  number={2-3},
  pages={93--100},
  year={2024},
  publisher={Wiley Online Library}
}

@article{miller2024far,
  title={How far are we off? Analyzing the accuracy of surgical margin relocation in the head and neck},
  author={Miller, Alexis and Wang, Vickie and Jegede, Victor and Necker, Fabian and Curry, Joseph and Baik, Fred M and Verma, Avanti and Holsinger, F Christopher and Tuluc, Madalina and Rahman, Mobeen and others},
  journal={Head \& Neck},
  year={2024},
  publisher={Wiley Online Library}
}

@article{pierik2021resection,
  title={Resection margins in head and neck cancer surgery: an update of residual disease and field cancerization},
  author={Pierik, Annouk S and Leemans, C Rene and Brakenhoff, Ruud H},
  journal={Cancers},
  volume={13},
  number={11},
  pages={2635},
  year={2021},
  publisher={MDPI}
}

@article{prasad2023augmented,
  title={Augmented-reality surgery to guide head and neck cancer re-resection: a feasibility and accuracy study},
  author={Prasad, Kavita and Miller, Alexis and Sharif, Kayvon and Colazo, Juan M and Ye, Wenda and Necker, Fabian and Baik, Fred and Lewis Jr, James S and Rosenthal, Eben and Wu, Jie Ying and others},
  journal={Annals of surgical oncology},
  volume={30},
  number={8},
  pages={4994--5000},
  year={2023},
  publisher={Springer}
}

@article{dzhugashvili2010surgical,
  title={Surgical clips assist in the visualization of the lumpectomy cavity in three-dimensional conformal accelerated partial-breast irradiation},
  author={Dzhugashvili, Maia and Pichenot, Charlotte and Dunant, Ariane and Balleyguier, Corinne and Delaloge, Suzette and Mathieu, Marie-Christine and Garbay, Jean-R{\'e}my and Marsiglia, Hugo and Bourgier, C{\'e}line},
  journal={International Journal of Radiation Oncology* Biology* Physics},
  volume={76},
  number={5},
  pages={1320--1324},
  year={2010},
  publisher={Elsevier}
}

@article{necker2023virtual,
  title={Virtual resection specimen interaction using augmented reality holograms to guide margin communication and flap sizing},
  author={Necker, Fabian N and Chang, Marcello and Leuze, Christoph and Topf, Michael C and Daniel, Bruce L and Baik, Fred M},
  journal={Otolaryngology--Head and Neck Surgery},
  volume={169},
  number={4},
  pages={1083--1085},
  year={2023},
  publisher={Wiley Online Library}
}

@article{barsouk2023epidemiology,
  title={Epidemiology, risk factors, and prevention of head and neck squamous cell carcinoma},
  author={Barsouk, Adam and Aluru, John Sukumar and Rawla, Prashanth and Saginala, Kalyan and Barsouk, Alexander},
  journal={Medical Sciences},
  volume={11},
  number={2},
  pages={42},
  year={2023},
  publisher={MDPI}
}

@article{van2019relocation,
  title={Relocation of inadequate resection margins in the wound bed during oral cavity oncological surgery: A feasibility study},
  author={van Lanschot, Cornelia GF and Mast, Hetty and Hardillo, Jose A and Monserez, Dominiek and Ten Hove, Ivo and Barroso, Elisa M and Cals, Froukje LJ and Smits, Roeland WH and van der Kamp, Martine F and Meeuwis, Cees A and others},
  journal={Head \& neck},
  volume={41},
  number={7},
  pages={2159--2166},
  year={2019},
  publisher={Wiley Online Library}
}

@article{heiselman2020intraoperative,
  title={Intraoperative correction of liver deformation using sparse surface and vascular features via linearized iterative boundary reconstruction},
  author={Heiselman, Jon S and Jarnagin, William R and Miga, Michael I},
  journal={IEEE transactions on medical imaging},
  volume={39},
  number={6},
  pages={2223--2234},
  year={2020},
  publisher={IEEE}
}

@inproceedings{RingelMorganandHeiselman2023RegularizedBreast,
  title={Regularized kelvinlet functions to model linear elasticity for image-to-physical registration of the breast},
  author={Ringel, Morgan and Heiselman, Jon and Richey, Winona and Meszoely, Ingrid and Miga, Michael},
  booktitle={International Conference on Medical Image Computing and Computer-Assisted Intervention},
  pages={344--353},
  year={2023},
  organization={Springer}
}

@inproceedings{ringel2024image,
  title={Image guidance system for breast conserving surgery with integrated stereo camera monitoring and deformable correction},
  author={Ringel, Morgan J and Richey, Winona L and Heiselman, Jon S and Stabile, Alexander and Meszoely, Ingrid M and Miga, Michael I},
  booktitle={Medical Imaging 2024: Image-Guided Procedures, Robotic Interventions, and Modeling},
  volume={12928},
  pages={59--67},
  year={2024},
  organization={SPIE}
}

@article{de2017regularized,
  title={Regularized kelvinlets: sculpting brushes based on fundamental solutions of elasticity},
  author={De Goes, Fernando and James, Doug L},
  journal={ACM Transactions on Graphics (TOG)},
  volume={36},
  number={4},
  pages={1--11},
  year={2017},
  publisher={ACM New York, NY, USA}
}

@article{perez2023ex,
  title={Ex vivo 3D scanning and specimen mapping in anatomic pathology},
  author={Perez, Alexander N and Sharif, Kayvon F and Guelfi, Erica and Li, Sophie and Miller, Alexis and Prasad, Kavita and Sinard, Robert J and Lewis Jr, James S and Topf, Michael C},
  journal={Journal of pathology informatics},
  volume={14},
  pages={100186},
  year={2023},
  publisher={Elsevier}
}

@article{teixeira2025outcomes,
  title={Outcomes in patients with head and neck squamous cell carcinoma with exclusively surgical resection},
  author={Teixeira, Daniel Naves Araujo and Lau, Fabio and de Oliveira, Vanessa Carvalho and Couto, Eduardo Vieira and Maahs, Thomas Peter and Lima, Carmen Silvia Passos and Chone, Carlos Takahiro},
  journal={Brazilian Journal of Otorhinolaryngology},
  volume={91},
  number={5},
  pages={101622},
  year={2025},
  publisher={Elsevier}
}

@article{nayanar2019frozen,
  title={Frozen section evaluation in head and neck oncosurgery: an initial experience in a tertiary cancer center.},
  author={Nayanar, Sangeetha K and KRISHNAN M, Aswathi and Mrudula, KI and THAVAROOL, P and Thiagarajan, Shivakumar and others},
  journal={Turkish Journal of Pathology},
  volume={35},
  number={1},
  year={2019}
}

@article{matsuo2024interval,
  title={Interval to recurrence affects survival in recurrent head and neck squamous cell carcinoma},
  author={Matsuo, Mioko and Hashimoto, Kazuki and Kogo, Ryunosuke and Sato, Masanobu and Manako, Tomomi and Nakagawa, Takashi},
  journal={Cancer Diagnosis \& Prognosis},
  volume={4},
  number={5},
  pages={658},
  year={2024}
}
\bibliographystyle{spiejour}   




\listoffigures
\listoftables

\end{spacing}
\end{document}